\documentclass{article}

\usepackage{amsmath}
\usepackage{amssymb}

\usepackage[colorlinks=true,allcolors=black]{hyperref}

\usepackage[tmargin=2cm,bmargin=2.5cm,lmargin=2.5cm,footnotesep=1cm]{geometry}

\usepackage[dvipsnames]{xcolor}

\newcommand{\rev}[1]{{#1}}

\begin{document}

\title{Machine learning and genomics: \\precision medicine vs. patient privacy}

\author{Chloe-Agathe Azencott\\
(1) MINES ParisTech, PSL Research University, CBIO--Centre for Computational Biology, \\F-75006 Paris, France \\
(2) Institut Curie, PSL Research University, F-75005 Paris, France\\
(3) INSERM, U900, F-75005 Paris, France\\
chloe-agathe.azencott@mines-paristech.fr}

\maketitle

\begin{abstract}
Machine learning can have major societal impact in computational biology applications. In particular, it plays a central role in the development of precision medicine, whereby treatment is tailored to the clinical or genetic features of the patient. However, these advances require collecting and sharing among researchers large amounts of genomic data, which generates much concern about privacy. Researchers, study participants and governing bodies should be aware of the ways in which the privacy of participants might be compromised, as well as of the large body of research on technical solutions to these issues. We review how breaches in patient privacy can occur, present recent developments in computational data protection, and discuss how they can be combined with legal and ethical perspectives to provide secure frameworks for genomic data sharing.
\end{abstract}

\section{Introduction}
The last decades have seen a surge in the amount and diversity of data collected to describe biological phenomena. 
This has opened the door to new approaches in biomedical research, allowing us to use statistical and machine learning approaches to objectively analyze observations and generate new hypotheses.

The benefits of these ``big data'' approaches range from advances in basic biology to computer-aided diagnosis. 
Among them, the promise of {\it precision medicine} -- tailoring care to the clinical, environmental and genetic characteristics of the patient -- is currently attracting a lot of interest, both in scientific communities and among the general public. 

Indeed, the differences in how patients experience the same disease, whether in terms of risk, prognosis, or response to treatment, can be striking. 
Even the most prescribed drugs for the most common conditions have very limited efficacy~\cite{schork2015}; however, this can be be explained in great part by genomic differences between patients~\cite{spear2001}. 
There is a lot of appeal, therefore, for data analysis solutions that allow us to exploit genomic \rev{databases} to precisely pinpoint these differences.

\subsection{Data-driven precision medicine}
Early examples of the usage of genomic information in precision medicine include the breast cancer drug trastuzumab (Herceptin), which dramatically improves the prognosis of patients whose tumor overexpresses the HER-2 gene, or the colon cancer drugs cetuximab (Erbitux) and panitumumab (Vectibix), which have little effect on patients that have a mutation in the KRAS gene.
The computational analysis of genomic data makes it possible to systematize such findings by 
finding genomic similarities among patients that exhibit the same prognosis, or response to a treatment.
However, while the number of potentially relevant genomic measurements is very large (for example, tens of thousands of proteins, hundreds of thousands of RNA transcripts, or millions of single point mutations), they are typically collected on only thousands of patients. 
In this setting, 
with many more variables than samples, statistical difficulties arise~\cite{donoho2000}. 
Data sharing, which is the most efficient way of increasing cohort sizes, and maximizes the utilization of each sample, is therefore crucial to the development of this field. 

It is also important to note that, because of the aforementioned challenges, existing approaches for the analysis of genomic data sets are limited in scope. There is a strong need to develop new statistical models and machine learning procedures to address these challenges, and hence to facilitate exploratory access to these data sets to computational biologists, statisticians and machine learners.\\

To illustrate our overview of privacy issues in genomic data sharing, we will focus here on {\it genome-wide association studies}, or GWAS. They are one of the most prominent \rev{tools} for detecting genetic variants correlated with an observed trait, such as disease status or response to treatment. 

They consist in collecting, for a large cohort of individuals, the variants they exhibit across hundreds of thousands to several millions of single nucleotide polymorphisms (SNPs), that is to say, individual locations across the DNA where variations of a single nucleotide can occur. 
A trait of interest (which can be binary, such as disease status, or continuous, such as age of onset) is also recorded for each individual. 
Statistical tests are then run to detect associations between the SNPs and the observed trait~\cite{gondro2013}. When the phenotype is binary, it is common to use $\chi^2$ or Cochran-Armitage tests, although they require the stringent assumptions that individuals come from a genetically homogeneous population, that the variants act independently from each other, and that no covariates such as age, sex, or environmental effects need to be accounted for. 
Relaxing these assumptions is an active area of statistical and machine learning research.


\subsection{Patient privacy}
While the potential scientific and social benefits of sharing genomic data are strong, these data are particularly sensitive. 

Indeed, the genomic sequence of an individual can be interpreted to discover information that most consider private, from their ancestry to their ability to metabolize certain drugs or the diseases they are more at risk of developing.
Moreover, genomic databases also hold sensitive clinical information, such as disease status, comorbidities, or environmental factors (which can include drug abuse or trauma).

In an era where data is sometimes touted as the new oil, there is a growing need to keep personal data private without impeding the technological, scientific and societal advances that will come from analyzing large data sets. This obviously applies to genomic data, even more so because of several specific features of these data. First, because our understanding of genetics is still growing, it is yet unclear what information the genome of an individual will reveal about them ten years from now. Second, genomic data does not only pertain to the person it belongs to: it also contains information, to some extent, about their family, or even about their ethnic, geographic or linguistic population. Finally, genomic data cannot be revoked: unlike a credit card that can be canceled once its number has been compromised, the genome of an individual cannot be changed.\\

And indeed, {\it genetic discrimination}, that is, being treated differently because you have, or are perceived to have, a particular genetic mutation, has been a source of concern for over thirty years~\cite{matthewman1984}. 
This concern was first addressed in the Declaration of Bilbao~\cite{bilbao1993}, which in 1993 denounced all uses of genetic information causing or leading to discrimination. 
Several laws have been passed across the world since then to implement such limitations. 
Currently, article 21 of the E.U. Charter of Fundamental Rights prohibits any discrimination based on, among other grounds\rev{, genetic features~\cite{article21}.}
In the US, the Genetic Information Non-discrimination Act (GINA)~\cite{gina} makes it illegal for employers or health insurers to use genetic test results. 
In several states, including California, GINA was later extended to other domains such as housing, mortgage lending, or education.
By contrast, the genetic test information of Canadians was not protected until January 2018~\cite{gna}.\\

In addition to laws regulating genetic discrimination, protection from breaches of genomic privacy can be achieved through {\it data protection regulations}. 
In the US, the Health Information Technology for Economic and Clinical Health (HITECH~\cite{hitech2009}) act requires data custodians to implement physical, administrative and technical solutions to appropriately protect biomedical data.
In the EU, the recently adopted General Data Protection Regulation (GDPR)~\cite{gdpr2016}\rev{, which will shortly come into force across Member States,} aims at preventing discriminatory effects on the basis of, among other causes, genetic or health status, and explicitly identifies genetic and biometric data as sensitive. Paired with this regulation, the Network and Information Security directive, which should be implemented in national laws by April 2018, includes health databases among the critical IT systems for which appropriate security measures to mitigate cybersecurity risks must be taken. \\

In spite of these policies, and of the lack of evidence for the occurrence of genetic discrimination, genetic privacy is still a cause for much concern across the world~\cite{green2015,wauters2016}. 
Indeed, the law does not cover all aspects of one's life that might be affected by genetic discrimination -- for example, social interactions. In addition, it is not because discrimination is illegal that it does not occur~\cite{quillian2017}.
It is therefore important to combine legal and ethical frameworks with technical solutions. 
In this paper, we will focus on the latest, and review the mathematical and computational tools that allow researchers to share genomic data without compromising patient privacy. 
We will start by reviewing the limitations of mere anonymisation. We will then describe two mathematical models for data privacy, the first one based on data suppression and the second on the addition of noise. Finally, we will outline encryption protocols and cryptographic hardware solutions.

\section{Anonymisation is not enough}
Genomic data, like any other sensitive data, is {\it anonymized} before being analyzed or shared. This means that unique identifiers such as names are stripped from it. Hence a patient who has taken part in a study about depression will not have her identity directly associated, in any database, with disease status (she has severe depression), comorbidities (she also has an anxiety disorder), environmental factors (she did drugs for ten years), or disease susceptibilities (she has a mutation that increases her risk for breast cancer).

However, anonymisation is not sufficient to guarantee unidentifiability. This is due to so-called {\it auxiliary information}: suppose you know that 53-year-old Jane was treated for colon cancer in 2016 at a given hospital. If your database for that hospital show\rev{s} only one record for a female patient of this age for this pathology, you will know this record must be Jane's.

This has made a number of attacks against genetic databases possible, such as using genealogies built from public records to identify participants in family-based studies~\cite{malin2006}, or assessing whether a given genotype is part of a cohort summed up by allele frequencies~\cite{homer2008,wang2009}. As a result, both the NIH and the Wellcome Trust updated their data sharing policy to strongly restrict access to individual genotypes and aggregate genotype frequency data~\cite{nih2007}. 

Many additional research results have shown the limitation of anonymisation. To give a few examples, 
\cite{schadt2012} showed how to predict the values of a thousand SNPs from easily available gene expression data, and how to re-identify individuals in SNP databases using these SNPs only; 
\cite{gymrek2013} identified people's surnames from profiling short tandem repeats on the Y chromosome and querying public genetic genealogy databases;
and \cite{claes2014} initiated work on predicting facial features from DNA sequences. More recently, the work of~\cite{lippert2017identification} on re-identification based on the combination of low-quality predictions of physical traits initiated a heated controversy~\cite{erlich2017,lippert2017no} on genomic data sharing.

In what follows, we will briefly review alternative approaches that provide appropriate participant privacy while maximizing scientific impact, or {\it utility.}

It should however be noted that genomes are the ultimate identifier: even if your name has been removed from a genomic database, your genome can be obtained from a drop of blood or some saliva. In this case, private information encoded in your DNA
could be obtained directly. In addition, an adversary could use your DNA to retrieve your record in a genomic database, hence gaining access to auxiliary information. This type of attack, however, requires physical access to a person's biological sample and to sequencing equipment. As such, it is much less scalable and, hence, worrying, than computer attacks based on database cross-referencing.

\section{Learning from obfuscated data}
\subsection{De-identification by data suppression}
Preventing re-identification often requires stripping more than unique identifiers from the data.
As an illustration, consider that an estimated $87\%$ of the U.S. population has a unique combination of date of birth, gender, and zip code~\cite{sweeney2000}. This made it possible to leverage voter registration data to find the names of $84$ to $97\%$ of the $579$ participants to the Personal Genome Project who had listed these three pieces of demographic information~\cite{sweeney2013}. Such pieces of information, which 
can be combined together to create a unique identifier,  are called {\it quasi-identifiers.}

In the US, the Health Insurance Portability and Accountability Act (HIPAA) therefore 
requires deleting not only names and phone numbers, but also additional information such as the final digits of all zip codes
, or vehicle identifiers. 

Unfortunately, some of the information that is destroyed can be crucial to a particular study. For example, when studying drug side effects, one can be interested in finding all cases where a rise in liver function tests followed the beginning of treatment by no more than a few days -- something that is not possible if dates have been rounded to the month of year. On the other hand, those guidelines are not sufficient to guarantee privacy in any database. 
Formal data protection models have been proposed by computer scientists to address these issues and quantify the notions of both utility and privacy. Among those, $k$-anonymisation, which we detail below, seeks to release the database, censored in such a way that re-identification is made almost impossible. 

\subsection{$k$-anonymisation}
{\it $k$-anonymity}, proposed by~\cite{sweeney2002}, seeks to prevent re-identification by stripping enough information from the released data that any individual record becomes indistinguishable from at least $(k-1)$ other records.   
For example, let us consider a database in which zip code and age are quasi-identifiers. While the three records \texttt{\{zip=47676, age=29, cancer=yes\}}, 
\texttt{\{zip=47677, age=26, cancer=no\}} and 
\texttt{\{zip=47272, age=27, cancer=yes\}} are all distinct, releasing them as 
\texttt{\{zip=47***, age=2*, cancer=yes\}}, 
\texttt{\{zip=47***, age=2*, cancer=no\}} and 
\texttt{\{zip=47***, age=2*, cancer=yes\}} ensures they all belong to the same equivalence class (i.e. the set of records that have the same identifying information) and satisfies $3$-anonymity.

However, $k$-anonymity does not provide any control on the variability of confidential but non-identifying attributes. In our example, it could very well be that the three records all have the confidential \texttt{cancer} attribute set to \texttt{yes}. In this case, a person matching any of these three records will be known to have cancer, even though their exact record will not be identifiable. 

Two variations on this framework, $l$-diversity~\cite{machanavajjhala2007} and $t$-closeness~\cite{li2007}, have therefore been proposed.
More specifically, a database release is said to satisfy {\it $l$-diversity} if any sensitive attribute has at least $l$ well-represented values in any equivalence class.
A database release is said to satisfy {\it $t$-closeness} if the distance between the distribution of a sensitive attribute within an equivalence class and its distribution within the entire database is upper-bounded by $t$. 

While $k$-anonymisation has been successfully applied to genomic sequences~\cite{li2012} or to electronic medical records used for the validation of GWAS~\cite{loukides2010}, 
it is important to note that this model and its variants offer no formal privacy guarantees. Indeed, 
because an adversary might have a greater level of knowledge than initially assumed, the above argument about auxiliary information still holds, and $k$-anonymisation is
vulnerable against several attacks~\cite{clifton2013}. One of these attacks, called a {\it linkage attack}, can be successful when another public database has information that overlaps with a $k$-anonymized data set. For example, the medical records of the governor of Massachussetts were identified from crossing anonymized medical data with public voter registration records.
In addition, when the number of sensitive variables is large, which is the case with genomic sequences and GWAS data, they can only be achieved by deleting most of these variables, hence losing in utility~\cite{aggarwal2005}.

\section{Learning from noisy data: differential privacy}
Another formal data protection model, called {\it differential privacy}~\cite{dwork2006,nissim2017}, was introduced to address the shortcomings of previous privacy models. In this popular model, which was awarded the G\"odel Prize in 2017, de-identification is prevented by the addition of noise to the data. The model is
\rev{based on the fact that auxiliary information will always make it easier to identify an individual in a dataset, even if anonymized.} 
Instead, differential privacy seeks to guarantee that the information that is released when querying a database is {\it nearly} the same whether a specific person is included in the study or not~\cite{dwork2006}. Unlike $k$-anonymity, differential privacy provides formal statistical privacy guarantees.

More specifically, a query function $\psi$ is called $\epsilon$-differentially private if, for any set $\mathcal{S}$ of possible answers, and for any two databases $\mathcal{D}$ and $\mathcal{D}'$ that differ by exactly one sample, the probability that the answer to $\psi$ on $\mathcal{D}$ is in $\mathcal{S}$ is $\epsilon$-close to the probability that the answer to the same query is in the same set when $\mathcal{D}'$ is queried: $\mathbb{P}(\psi(\mathcal{D}) \in \mathcal{S}) 
 \leq e^\epsilon \; {\mathbb{P}(\psi(\mathcal{D}') \in \mathcal{S})}.$
 
Both an interactive and a non-interactive settings can be considered. In a non-interactive setting, an obfuscated version of the database is released. Alternatively, only summary statistics are released, possibly after the addition of noise. In an interactive setting, the database is not made public, but users are allowed to query it, and it is the true answers to these queries that are obfuscated by the injection of noise.

This noise injection obviously leads to somewhat inaccurate results.
The goal of differential privacy research is therefore to simultaneously maximize utility (i.e. limit the inaccuracy of the results) and privacy (i.e. minimize $\epsilon$).
While $k$-anonymity and differential privacy are often opposed, 
the latter having formal privacy guarantees but often lower utility than the former,
efforts are being made to bridge the gap between both privacy models~\cite{soriacomas2014}.

Unfortunately, in precision medicine applications, the dimensionality of the data tends to be a major limitation to the utility of differentially private mechanisms
~\cite{fredrikson2014}
.
Nevertheless, recent work~\cite{johnson2013,yu2014b,simmons2016} has shown how to implement differential privacy for GWAS, making it possible for users to query a genetic database for the differentially private top $k$ most associated SNPs. The results are approximate in the sense that the returned $p$-values are exact only within an order of magnitude, and the exact location of the SNPs might be off; however, these results are believed to be of sufficient quality to drive further biomedical analyses. 

A major drawback of these approaches, however, is that only a preset list of queries (``return the $p$-value of a given SNP'', ``return the location of the top $k$ SNPs'') are allowed. This restricts the diversity of exploratory analyses one might conduct on such data. 
In a field where statisticians and machine learners are still proposing novel methods to get more informative results from data, this is a strong limitation. There is still a strong need for mechanisms for releasing a noisy version of the data, on which one can efficiently get highly accurate answers from a variety of queries~\cite{dankar2012}. 

\section{Working with encrypted genomic data}
In the interactive differential privacy setup above, the data is only visible to users as the distorted outcomes of authorized queries. However, the data server itself is not immune to attacks. 
In addition, differential privacy solutions are limited in that they result in noisy outcomes. 
There is therefore a growing interest for using cryptographic protocols to share and analyze biomedical data without revealing the contents of any particular record. This goal is achieved while providing formal computational privacy guarantees, which depends on how much computation is allowed on the server on which the private data resides. The yearly competitions organized by the NIH-funded national center for integrating data for analysis, anonymisation and sharing (iDASH\footnote{\url{http://www.humangenomeprivacy.org}}) are instrumental in the development of privacy by encryption.

\subsection{Homomorphic encryption}
As cohort sizes grow, so does the interest for performing genomic analyses on the cloud (as demonstrated, for example, by the Pan-Cancer Analysis of Whole Genomes project~\cite{molnargabor2017}). 
Reliable technical solutions for securely outsourcing genomic data analyses to remote servers are therefore necessary.
{\it Homomorphic encryption} is such a solution. 
It seeks to ensure that computation on encrypted data yields a result which, once decrypted, matches the result you would have gotten on the non-encrypted data. This makes it possible to entrust a third party with the encrypted version of your data, knowing that they will not be able to decrypt it, nor the results of your analyses. 

Such a scheme could also be used to facilitate sharing genomic data that are stored encrypted in the cloud. Arbitrary individuals can then query the encrypted data; however, their answer is only decrypted (by an authority) if their queries match a predefined allowable query policy~\cite{lauter2015}. 

Unfortunately, homomorphic encryption techniques only support a small number of arithmetic operations on the data, and are demanding both in computational time and in memory. Still, they have been successfully applied to searching sequences of SNPs in large genomic databases~\cite{shimizu2016}. It is also possible to compute $\chi^2$ association tests for an entire GWAS data set over a population of $10\,000$ patients in half a day~\cite{lu2015}, although the storage costs of key generations can be high~\cite{zhang2015foresee,kim2015}. A solution for rare variants analysis, by means of an exact logistic regression model, has also been proposed~\cite{wang2016}.
The computational and memory costs of homomorphic encryption schemes are still several orders of magnitude larger than the equivalent computation over non-encrypted data. In addition, the solutions that have been proposed are tailored to a few specific analyses, and are far from covering all existing methods for the analysis of GWAS data. They are therefore, at the moment, of limited interest to the practitioner.

\subsection{Secure multi-party computation}
A common data sharing setup is that of collaborative analyses, in which multiple institutions would like to join their data sets and all obtain the results from queries on the combined data set. 
Setting up a platform where genomic data can be stored and processed so that none of the individual institutions can reconstruct the data reduces the risk of both inadvertent and malicious leaks, hereby facilitating such a collaboration.
This is the problem that the field of {\it secure multi-party computation} addresses.
This subfield of cryptography aims at providing methods for multiple parties to jointly compute a function over the union of their input data while keeping those inputs private. \\

Secure multi-party computations were first introduced by Yao~\cite{yao1986} for two parties. Assuming two agents $A$ and $B$ each own data $\mathcal{D}_A$ and $\mathcal{D}_B$, the goal is for both $A$ and $B$ to be able to compute a function $f(\mathcal{D}_A, \mathcal{D}_B)$ without either of them learning the data from the other. In this setting, the main approach is known as {\it garbled circuits}, or {\it Yao's protocol}. It requires encoding $f$ as a Boolean function, or circuit, with fixed-length binary inputs. It is possible for $A$ to randomize (``garble'') the truth tables of the logical gates of $f$ into $\hat f$, as well as her input $\mathcal{D}_A$ into $\widehat{\mathcal{D}_A}$, in such a way that $\hat f(\widehat{\mathcal{D}_A}, \widehat{\mathcal{D}_B}) = f(\mathcal{D}_A, \mathcal{D}_B)$. $A$ transmits $\hat f$ and $\widehat{\mathcal{D}_A}$ to $B$, who also obtains $\widehat{\mathcal{D}_B}$ from $A$ using {\it oblivious transmission}, meaning that $A$ never learns $\mathcal{D}_B$ in the process. $B$ can then reveal the output of $\hat f(\widehat{\mathcal{D}_A}, \widehat{\mathcal{D}_B})$ to $A$.

Multiple garbled circuit methods have been proposed to analyze genomic data, in particular for computing similarities between sequences~\cite{jha2008,wang2015} or for case-control GWAS studies~\cite{constable2015}. However, these approaches are limited to two-party computations, meaning that they yet have to be adapted to the case where more than two partners want to collaborate on a federated genomic study. \\

In contrast, {\it secret sharing} approaches to secure multi-party computations guarantee that the database can only be reconstructed if $t$ out of the $n$ participants collude together.
Here, all parties have a similar role (unlike the sender $A$ and the receiver $B$ in garbled circuits), and the functions are encoded are {\it arithmetic circuits}, meaning that instead of logical gates, the circuits are composed of additions and multiplications.

Although current secret sharing solutions have strong limitations (for example, floating-point and comparison operations are inefficient), they have been successfully applied to GWAS data, both for association tests~\cite{kamm2013, zhang2015secure} and minor allele frequency computations~\cite{zhang2015secure}.
In addition, \cite{xie2014} showed how to account for differences in study design between the different institutes.

Although they apply to a different setting, these approaches are more efficient than homomorphic encryption schemes. Unfortunately, they still have limited flexibility, as specific circuits must be designed and optimized for each task. 
In addition, they do not allow additional users to perform computations on the data, which strongly limits their application to genomic research.

\subsection{Cryptographic hardware}
Both homomorphic encryption and secure multi-party computations impose significant computational overhead when compared to analyses over non-encrypted data. In addition, the type of analyses that can be implemented within these frameworks is limited. 
By contrast, solutions based on cryptographic hardware do not significantly increase computational time, nor do they severely limit the operations that can be executed on the data. They are therefore an interesting avenue to make secure large-scale genetic analyses feasible in practice.

{\it Secure co-processors}, or {\it hardware security modules}, are computational devices that can be trusted to store data execute code securely, even against an adversary who physically controls the host.
A first example or their use in genomics data processing is~\cite{canim2012}, in which secure co-processors were used to securely query a genomic database for the number of samples matching a given SNP pattern.

A more popular architecture is Intel {\it 
Software Guard Extensions (SGX)}~\cite{anati2013}. It extends  the architecture of Intel x86 processor and allows for the creation of private memory regions, in which code and data are isolated and protected from external processes. This architecture makes it possible to allow untrusted parties to remotely run computations on private data without compromising them. SGX computations are several orders of magnitudes more efficient than homomorphic encryption or secure multi-party computations.
For example, \cite{chen2017princess} leverages SGX to compute global transmission/disequilibrium test statistics from allele counts transmitted by individual research centers having each performed family studies, therefore enabling a real-world collaboration on Kawasaki's disease.

\subsection{Protecting genomic databases}
Data obfuscation and differential privacy aim at ensuring sensitive data is not revealed from inferences drawn from the data that is released. Genomic data encryption
safeguards privacy in that the data that is stored is encrypted; even in the case of direct attacks against the databases, keys are needed to decrypt the data. Finally, computer security aims at ensuring that only people with authority to receive information have access to it. 
\rev{While neither computer security nor genomic data encryption}
deal with inferences that can be drawn from answers to authorized queries, it is an important aspect of safeguarding genomic data. Indeed, the HITECH act in the US and the GDPR~\cite{gdpr2016} in Europe are now requiring that genomic data custodians implement physical, administrative and technical solutions to appropriately protect biomedical data.

Encryption techniques such as homomorphic encryption, secure multi-party computation, and the use of cryptographic hardware go a long way towards achieving these goals. In addition, a growing interest for the {\it blockchain} technology has driven a number of commercial initiatives~\cite{blockchain2017, grishin2018}. It is important to note, however, that, on its own, a blockchain can only be used to guarantee that only authorized actors can download a database. In other words, these actors obtain access to the raw data. To increase both privacy and utility, it will be desirable to limit the number of users who have access to the raw data, while increasing the number of users who can query it.
 This requires combining the blockchain mechanism with additional cryptographic or differential privacy techniques.

\section{Discussion}
Protecting the privacy of 
all individuals who have their genome sequenced, or other types of genomic data measured,
is a concern of growing importance. 
There is at the moment little evidence of actual cases of ``genomic hacking'', outside of the somewhat artificial conditions of re-identification experiments. However, as genomic data become technically easier to acquire, and as our ability to interpret them grows, it is not outlandish to expect more occurrences of private genomic information leaks or genetic discrimination.
Nevertheless, while genomic data are highly sensitive, they are also, thanks to machine learning and statistics, key to the progress of healthcare and precision medicine. 
It is therefore important to accumulate such data and devise ways for researchers to access them securely.
To guarantee that the scientific and social benefits of genomic data sharing outweigh the potential pitfalls, regulators and scientists must work together to develop appropriate sharing frameworks, based on ethical concerns and technical solutions.\\

In agreement with~\cite{murphy2011}, we propose to build our reflection around three aspects.
The first is {\it algorithmic solutions to de-identification}, that is to say data (or query answer) obfuscation solutions. As we have seen, $k$-anonymity and its derivatives are currently limited in the level of privacy they can effectively guarantee for patients. Differential privacy, which returns noisy answers to database queries, is a powerful alternative; however its utility is limited by the need to return somewhat inaccurate answers.
In addition, the scope of queries that can be performed is narrow, which currently limits the development of novel machine learning methods in the context of differential privacy.

The second is {\it database security}. Cryptographic frameworks such as homomorphic encryption or secure multi-party computation allow authorized users to perform specific queries without ever accessing the data. In addition, because the data is stored encrypted, it is also protected against direct attacks. While these frameworks only allow limited types of requests at the cost of significant computational and storage overheads, cryptographic hardware is emerging as a much efficient way to perform secure and accurate analyses.

\rev{It is important to note that differential privacy and cryptography are complementary techniques: in the first one, the guarantee of privacy is statistical, and is degraded when the number of data queries increases, while in the second, the guarantee of privacy is computational, and is degraded when the server is allowed unbounded computations.}
Both approaches to data protection are still in their infancy, particularly pertaining to the specific challenges of genomic applications. Technical developments in these fields should be pushed forwards to address the limitations we highlighted. 

Finally, because unlike data obfuscation solutions, which are meant to let any user access or query the data, cryptographic solutions only let authorized users query the database, the third important front to consider is {\it researcher trust}: who are the authorized users? Indeed, cryptographic solutions are designed to only let data owners analyze their data, or a larger collection of data they have contributed to. 
On the one hand, this circumvents the significant resources associated with setting up a granting authority who verifies whether researchers asking for data access are legitimate.
On the other hand, 
restricting the number of well-meaning researchers that can access these databases strongly limits their potential impact.
Data access restrictions are a burden for researchers, particularly junior researchers or small labs that do not have the clout to set up collaborations with major data curators.\\

Several interesting ideas regarding trust have been discussed in recent years. For example,~\cite{wan2017} have demonstrated how to use game theory to devise optimal policies to choose between (1) partial but unconditional data sharing and (2) data usage agreements, complete with financial penalties in the event of a breach of contract. 

Another approach is to somewhat give up on the notion of privacy, by informing study participants that, although appropriate measures will be taken to that effect, their privacy cannot be guaranteed~\cite{lunshof2008, ball2012}. Hence the ``trust not privacy'' approach, which fits well in the emerging paradigm of so-called P4 (preventive, predictive, personalized and participatory) medicine, aims at alleviating the concerns participants may have by involving them in the research process. In essence, the knowledge that they are advancing science compensates the potential drawbacks of their volunteerism. This can be achieved by informing participants of the intended and actual use of their data; giving them control over future use of their data; and informing them of the outcomes of the studies they have participated in~\cite{erlich2014}. One could also contemplate combining this vision with specific insurance policies against privacy attacks for genomic studies participants.\\

It is important to point out that the issues we have discussed are not limited to academic studies of genetic data.
On the contrary, direct-to-consumer genetics testing companies accumulate large amounts of genetic data, raising
potentially problematic issues in terms of the adequacy of the consumer's informed consent~\cite{niemiec2016}, in particular with respect to privacy issues.
In addition, although we have focused here on genomic data, increasing amounts of non-genomic biomedical data are being accumulated from wearable medical devices, fitness tracking devices, at-home sensors; and these potentially sensitive data are not necessarily covered by existing legislation, nor the focus of current privacy and cryptography research efforts.\\

To conclude, it is important to stress that the mechanisms of gene expression and its regulation are complex. DNA alone does not determine behavior, intelligence, or health. When it comes to complex traits, most effect sizes we observe are very small, meaning that they can only be interpreted as small risk increases. We must also remember that, even with advanced machine learning techniques, there is a long way to go from statistical observations to molecular mechanisms. In addition, current cohorts are predominantly white, male, from developed countries; whether the findings apply to individuals from other populations is unclear. This does not lower the value of genetic studies for biomedical advances, as there is still a lot left to understand about the role of our genome in diseases and responses to treatment. However this means that potential actors in genetic discrimination should be educated about the limited extent of the information one can learn about someone from their genomic data. It is often the accompanying clinical data, such as disease status, comorbidities or environmental factors, that can be the most damaging when used wrongly.

\enlargethispage{20pt}




The author declare that she has no competing interests.


The author would like to thank participants to discussions ensuing her talk on a similar topic at the Machine Learning in Society Workshop at DALI 2016 in Sestri Levante, Italy on April 1st, 2016.



\end{document}